\documentstyle[prd,aps,multicol,epsf]{revtex}
\renewcommand{\narrowtext}{\begin{multicols}{2}
\global\columnwidth20.5pc}
\renewcommand{\widetext}{\end{multicols} \global\columnwidth42.5pc}
\def\Lrule{\vspace*{-0.2in}\noindent\vrule width3.5in height.2pt
  depth.2pt \vrule depth0em height1em}
\def\Rrule{\vspace{-0.1in}\hfill\vrule depth1em height0pt \vrule
  width3.5in height.2pt depth.2pt\vspace*{-0.1in}}
\def\bml{\begin{mathletters}}
\def\eml{\end{mathletters}}
\def\beq{\begin{equation}}
\def\eeq{\end{equation}}
\def\bea{\begin{eqnarray}}
\def\eea{\end{eqnarray}}
\def\ba{\begin{array}}
\def\ea{\end{array}}
\def\noi{\noindent}
\def\nn{\nonumber}
\def\to{\rightarrow}

\def\a{\alpha}

\def\m{\mu}

\def\z{\zeta}
\def\tz{\tilde{\zeta}}
\def\tm{\tilde{\mu}}

\def\tm{\tilde{\mu}}
\def\e{{\rm e}}
\def\tr{\,{\rm tr}\,}
\begin{document}
\preprint{TIT-HEP-446}
\draft
\title{
Distribution of the k-th smallest Dirac operator eigenvalue
}
\author{Poul H. Damgaard}
\address{
The Niels Bohr Institute, Blegdamsvej 17, DK-2100 Copenhagen \O,
Denmark 
}
\author{Shinsuke M. Nishigaki}
\address{
Department of Physics,
Tokyo Institute of Technology,
O-okayama, Meguro, Tokyo 152-8551, Japan
}
\date{June 15, 2000;\ revised July 14, 2003}
\maketitle
\begin{abstract} 
Based on the exact relationship to
Random Matrix Theory, we derive
the probability distribution of the $k$-th smallest Dirac operator
eigenvalue in the microscopic finite-volume scaling regime of QCD
and related gauge theories.
\end{abstract}
\renewcommand{\thefootnote}{\fnsymbol{footnote}}
\setcounter{footnote}{0}
\narrowtext
While index theorems have for long been known to relate the number of
exact zero modes of the Dirac operator to
gauge field topology, it is only during the 1990's that it has been
realized how much further one can push such exact predictions. When
the gauge theory in question is in a phase with spontaneously broken
chiral symmetry, Goldstone's Theorem allows for much more detailed
statements, beyond the zero modes. The bridge between the
Dirac operator spectrum and the Goldstone degrees of freedom is the
effective partition function of the associated chiral Lagrangian
\cite{LS,SmV}. The infinite-volume chiral condensate will be denoted
by $\Sigma$.
By considering the theory at finite four-volume $V$, one
can choose a {\em microscopic scaling regime} in which this volume is sent
to infinity at a rate correlated with the chiral limit
$m \to 0$ such that $\mu\equiv \Sigma V m$ is kept fixed.
Then exact statements can be made for an infinite sequence of Dirac
operator eigenvalues that accumulate towards the origin. This realization
was first made on the basis of a universal Random Matrix Theory
reformulation of the problem \cite{ShV,V,HV,VW},
but it has since then
also been established directly at the level of the effective Lagrangian
\cite{D,OTV}. 

Conventionally, the exact analytical statements that can be made about
this infrared region of the Dirac operator spectrum in finite-volume
gauge theories have been phrased in terms of the microscopic spectral
$n$-point functions, and in particular the microscopic spectral density
$\rho_S(\zeta;\{\mu\})$ itself (here $\zeta \equiv \Sigma V \lambda $
is the microscopically rescaled Dirac operator eigenvalue $\lambda$)
\cite{ShV,V,NS,For1,For2,AST,NF95,DN98a,WGW,NN1,AK,NN2}.
Also the exact distribution of just the {\em lowest} Dirac operator
eigenvalue has been derived on the basis of the relationship to Random
Matrix Theory \cite{For1,SlN,TW,FH,NF98,NDW,WGW,NN1}.
In fact also the distribution
of the second-smallest eigenvalue in what corresponds to the massless limit
of $SU(N_c\geq 3)$ gauge theories with fermions in the fundamental
representation can be extracted from the work of Forrester and Hughes
\cite{FH}. In this paper, we
push these computations one step further
and provide analytical expressions for
both the joint probability distribution of the first $k$ eigenvalues
and the distribution of the $k$-th eigenvalue. We do so
for the chiral unitary ($\beta=2$), orthogonal ($\beta=1$,
corresponding to $SU(2)$ gauge theories with fundamental fermions), and
symplectic ensembles
($\beta=4$, corresponding to
$SU(N_c\geq 2)$ gauge theories with adjoint fermions). We will treat
the most general case of massive fermions, and (with one exception noted
below) a sector of arbitrary topological charge $\nu$.
In the microscopic
scaling limit this can be done to arbitrarily high order $k$ as long as
the volume $V$ is taken large enough.
This gives us an infinite sequence of
distributions of Dirac operator eigenvalues that all, on the macroscopic
scale, build up the spectral density at the origin $\rho(0)$. We
stress
that there is much more information in these individual eigenvalue
distributions than in the summed-over microscopic spectral density
$\rho_S(\zeta;\{\mu\})$ itself. In particular, from the point of view of
lattice gauge theory simulations it is far more convenient to be able to
perform comparisons with individual eigenvalue distributions than with just
the average eigenvalue density.

There are by now detailed analytical predictions for the microscopic
spectral densities of the Dirac operator for all three universality
classes.
These microscopic spectral densities show
a typical oscillatory structure, which clearly is closely connected
to the individual eigenvalue peaks at the microscopic scale where units
are given by the average eigenvalue spacing. We thus expect that the
distribution of the $k$-th smallest Dirac operator eigenvalue corresponds
closely to the $k$-th peak in the microscopic eigenvalue density, a
feature that indeed had already been noticed in the case of just the
smallest eigenvalue \cite{NDW}. In particular, we expect that the
$k$-th eigenvalue distribution as computed in the framework of Random
Matrix Theory is universal, which indeed turns out to be the case (see
below). Of course, as one further additional check,
the sequential summing-up of the individual eigenvalue distributions
should simply build up the full microscopic eigenvalue densities, which
are now known in closed analytical form.

The chiral Random Matrix Theory ensembles
are defined by \cite{ShV,V,HV}:
\widetext
\beq
{Z}^{(\beta)}_\nu (m_1,\ldots,m_{N_f})
=\int dW \e^{- \beta \tr v(W^\dagger W)}
\prod_{i=1}^{N_f}
\det 
\left(
\ba{cc}
m_i & W \\
-W^\dagger & m_i
\ea
\right),
\label{ZchiRME}
\eeq
where the integrals are over complex, real, and quaternion real
$(N+\nu)\times N$ matrices $W$ for $\beta=2,1,4$, respectively.
These chiral random matrix ensembles provide exactly equivalent
descriptions of the effective field theory partition functions in
the microscopic finite-volume scaling regime \cite{ShV,HV}. Moreover,
all microscopic spectral properties of the Dirac operator coincide
exactly with the corresponding microscopically rescaled Random
Matrix Theory eigenvalues \cite{OTV}. This means that either formulation
can be used to derive identical results, and in the case of the
individual distributions of the smallest Dirac operator eigenvalues
it is presently more simple to use the Random Matrix Theory formulation.
Since the results turn out to be universal, $i.e.$ independent of the
detailed form of the Random Matrix Theory potential $v(W^\dagger W)$
\cite{ADMN,KF,SeV},
it suffices for us to concentrate on Gaussian
ensembles with $v(x)= x$.
This choice leads to Wigner's semi-circle law
$\rho(\lambda)=(2/\pi)\sqrt{2N-\lambda^2}$.

The partition functions (\ref{ZchiRME})
of such chiral Gaussian ensembles
corresponding to $N_f$ massive flavors and topological charge $\nu$
can then be written in terms of eigenvalues $\{x_i\}$ of the
positive-definite Wishart matrices $W^\dagger W$,
\beq
Z^{(\beta)}_\nu (m_1,\ldots,m_{N_f})=
\Bigl( \prod_{i=1}^{N_f} m_i \Bigr)^\nu
\int_0^\infty \!\!\!\cdots \int_0^\infty
\prod_{i=1}^N 
\Bigl(dx_i\, x_i^{\beta(\nu+1)/2-1}
\e^{-\beta x_i}
\prod_{j=1}^{N_f} (x_i+m_j^2) \Bigr)
\prod_{i > j}^N |x_i-x_j|^\beta ~, \label{Zeigenv}
\eeq
up to an overall irrelevant factor which is independent of the $m$'s.
Because everything will be symmetric under $\nu \to -\nu$, we for
convenience take $\nu$ to be non-negative.
In the following, we impose the one single technical restriction that
for $\beta=1$
the topological charge $\nu$ be odd.

The unnormalized 
joint probability distributions
for $N$ eigenvalues $\{0\leq x_1\leq \cdots \leq x_N\}$
in the above Random Matrix Theory ensembles take the form
\beq
\rho^{(\beta)}_N (x_1,\ldots,x_N;\{m^2\})=
\prod_{i=1}^N \Bigl(x_i^{\beta(\nu+1)/2-1}
\e^{-\beta x_i} \prod_{j=1}^{N_{f}} (x_i+m^2_j) \Bigr)
\prod_{i > j}^N |x_i-x_j|^\beta ~.
\eeq
The joint probability distribution of
the $k$ smallest eigenvalues
$\{0\leq x_1\leq \cdots \leq x_{k-1}\leq x_k\}$
can be written as
(the order of $x_1,\ldots,x_{k-1}$ can be relaxed):
\bea
&&{\Omega}^{(\beta)}_{N,k}(x_1,\ldots,x_{k-1},x_k;\{m^2\})
\equiv \frac{1}{\Xi^{(\beta)}_{N}(\{m^2\})}
\frac{1}{(N-k)!}
\int_{x_{k}}^\infty dx_{k+1} \cdots
\int_{x_{k}}^\infty dx_{N}\,
\rho^{(\beta)}_N (x_1,\ldots,x_N;\{m^2\})
\nonumber\\
&&=\frac{1}{\Xi_N^{(\beta)}(\{m^2\})}
\prod_{i=1}^k \Bigl(x_i^{\beta(\nu+1)/2-1} \e^{-\beta x_i}
\prod_{j=1}^{N_{f}} (x_i+m^2_j) \Bigr)
\prod_{i>j}^k |x_i-x_j|^\beta
\nonumber\\
&&\times\frac{1}{(N-k)!}
\int_{x_{k}}^\infty dx_{k+1} \cdots
\int_{x_{k}}^\infty dx_{N}\,
\prod_{i=k+1}^N \Bigl(x_i^{\beta(\nu+1)/2-1} \e^{-\beta x_i}
\prod_{j=1}^{N_{f}} (x_i+m^2_j)
\prod_{j=1}^{k} (x_i-x_j)^\beta \Bigr)
\prod_{i >j\geq k+1}^N |x_i-x_j|^\beta ~,
\label{Omega}
\eea
where $\Xi^{(\beta)}_{N}$ stands for
the normalizing integral
\beq
\Xi^{(\beta)}_{N}(\{m^2\})=\frac{1}{N!}
\int_0^\infty \!dx_1 \cdots\int_0^\infty \!dx_N\,
\rho^{(\beta)}_N (x_1,\ldots,x_N;\{m^2\})~ .
\eeq

We now shift $x_i \to x_i + x_k$ in the integrand of (\ref{Omega}):
\bea
&&{\Omega}^{(\beta)}_{N,k}(x_1,\ldots,x_{k-1},x_k;\{m^2\})
~=~ \frac{e^{-(N-k)\beta x_k}}{\Xi_N^{(\beta)}(\{m^2\})}\frac{1}{(N-k)!}
\prod_{i=1}^k \Bigl(x_i^{\beta(\nu+1)/2-1} \e^{-\beta x_i}
\prod_{j=1}^{N_{f}} (x_i+m^2_j) \Bigr)
\prod_{i>j}^k |x_i-x_j|^\beta \cr
&&\times \int_{0}^\infty \prod_{i=k+1}^N \Bigl(dx_{i}\, \e^{-\beta x_i}
x_i^{\beta}(x_i + x_k)^{\beta(\nu+1)/2-1}
\prod_{j=1}^{N_{f}} (x_i+m^2_j+x_k) \prod_{j=1}^{k-1}(x_i + x_k -
x_j)^{\beta}
\Bigr)
\prod_{i>j\geq k+1}^N |x_i-x_j|^{\beta}~. 
\label{shifted}
\eea
To get the probability distributions of the Dirac operator eigenvalues
we must change the picture back
to chiral Gaussian ensembles, and take the
microscopic limit
\beq
\ba{l}
x_i \to 0 \\
m^2_j \to 0
\ea ,
\ \ N\to \infty, \ \ \mbox{with the rescaled variables}~\ \
\ba{ll}
\zeta_i=\pi \rho(0) \sqrt{x_i} &=\sqrt{8N\, x_i} \\
\mu_j=\pi \rho(0) m_j  &=\sqrt{8N}m_j
\ea  ~~~{\rm kept ~fixed} .
\label{microlim}
\eeq
In this large-$N$ limit the difference between partition functions
based on $N-k$ and $N$ eigenvalues becomes insignificant. Moreover,
one notices that the new terms in the integrand of (\ref{shifted}) can
be interpreted as arising from new additional fermion species, with
the partition function now being evaluated in a fixed topological sector of
effective charge $\nu = 1 + 2/\beta$. Taking into
account the definition (\ref{Zeigenv}), we then get, with 
${\cal Z}^{(\beta)}_{\nu}(\{\mu\})$ denoting 
the partition functions in the microscopic limit,
\bea
&&\omega_k^{(\beta)}(\zeta_1,\ldots,\zeta_k; \{\m\})
~=~ \lim_{N\to \infty} \Bigl(\prod_{i=1}^k \frac{|\zeta_i|}{8N} \Bigr)
{\Omega}_{N,k}^{(\beta)}
(\frac{\zeta_1^2}{8N},\ldots,\frac{\zeta_k^2}{8N};
\{\frac{\m^2}{8N} \})
\nn\\
&&=
{\rm const.}\,
\e^{-\beta \zeta_k^2/8}
\zeta_k^{\beta\frac{\nu+1}{2}-\nu-1+\frac{2}{\beta}}\prod_{j=1}^{N_{f}}
(\mu_j^2+\zeta_k^2)^{\frac{1}{2} - \frac{1}{\beta}}
\prod_{i=1}^{k-1}\Bigl(\zeta_i^{\beta(\nu+1)-1}
(\zeta_k^2-\zeta_i^2)^{\frac{\beta}{2}-1}
\prod_{j=1}^{N_{f}}(\zeta_i^2+\mu_j^2)
\Bigr)\prod_{i>j}^{k-1}(\zeta_i^2-\zeta_j^2)^{\beta}
\prod_{j=1}^{N_{f}}\mu_j^{\nu} \cr
&&\times
\frac{{\cal Z}^{(\beta)}_{1+2/\beta}(
\sqrt{\mu_1^2+\zeta_k^2},\ldots,\sqrt{\mu_{N_{f}}^2+\zeta_k^2},
\overbrace{\sqrt{\z_k^2-\z_1^2},\ldots,\sqrt{\z_k^2-\z_1^2}}^{\beta},
\ldots,
\overbrace{\sqrt{\z_k^2-\z_{k-1}^2},\ldots,\sqrt{\z_k^2-\z_{k-1}^2}}^{\beta}
,
\overbrace{\zeta_k,\ldots,\zeta_k}^{\beta(\nu+1)/2-1})}{
{\cal Z}^{(\beta)}_{\nu}(\mu_1,\ldots,\mu_{N_{f}})} ~.
\label{bboxOmega}
\eea
This is the main result of our paper. It shows that the joint probability
distribution of the first $k$ eigenvalues is given, apart from the
relatively simple prefactor, by a ratio of
finite-volume partition functions.
The new partition function that enters in the
numerator of Eq.(\ref{bboxOmega}) has the $N_f$
original fermion masses shifted according to $\mu_i \to
\sqrt{\mu_i^2+\zeta_k^2}$, contains $\beta(k-1)$ new fermions
of masses $\sqrt{\z_k^2-\z_1^2}, \ldots, \sqrt{\z_k^2-\z_{k-1}^2}$
(each mass $\beta$-fold degenerate), $\beta(\nu+1)/2-1$
additional degenerate fermion species of common mass $\zeta_k$,
and this whole partition function is evaluated in a sector of fixed
topological charge $\nu=1+2/\beta$. While this topological index
is fractional for $\beta=4$, there is no difficulty with the evaluation
of the pertinent partition function. (Indeed, it can alternatively
be viewed as a partition function in a sector of topological charge
$\nu=0$ and 3 additional massless fermions.) This expression
entirely in terms of {\em the effective field theory} partition
functions strongly suggests that it should
be possible to derive these analytical expressions starting directly
from the effective field theory, perhaps partially quenched as in
Ref.\cite{OTV}. The proportionality 
constant in Eq.(\ref{bboxOmega})
depends on the normalization conventions
of the involved partition functions. We fix this numerical factor
uniquely by the requirement that the total probability of finding any
given eigenvalue is normalized to unity.

Correlation functions of the rescaled
eigenvalues $\{\zeta_{k_1},\ldots, \zeta_{k_{p-1}}, \zeta_k\}$
are obtained from 
$\omega^{(\beta)}_{k}(\zeta_1,\ldots,\zeta_k;\{\m\})$
by integrating out the remaining eigenvalues
in a cell $0\leq \zeta_1\leq \cdots \leq \zeta_k$.
In particular, the distribution
of the $k$-th smallest eigenvalue $\zeta$,
${p}^{(\beta)}_{k}(\zeta;\{\mu\})$,
is given by 
\beq
{p}^{(\beta)}_{k}(\zeta;\{\mu\})=
\int_0^\zeta d\zeta_1 \int_{\zeta_{1}}^\zeta d\zeta_2 
\cdots \int_{\zeta_{k-2}}^\zeta d\zeta_{k-1}\,
\omega^{(\beta)}_{k}(\zeta_1,\ldots,\zeta_{k-1},\zeta;\{\m\})~ .
\eeq
We note that this gives a very simple representation of the probability
distribution of the $k$-th smallest Dirac operator eigenvalue. In
particular,
already for the most elementary case of just the {\em lowest} Dirac operator
eigenvalue in the $\beta=2$ universality class we obtain a very simple 
expression for an arbitrary
topological sector $\nu$. In the normalization convention of, $e.g.$,
\cite{D},
\beq
p_1^{(2)}(\zeta,\{\mu\}) =  \frac{\zeta}{2}\e^{-\zeta^2/4}
\Bigl(\prod_{j=1}^{N_{f}}\mu_j\Bigr)^{\nu}
\frac{{\cal Z}_2(\sqrt{\mu_1^2+\zeta^2},\ldots,
\sqrt{\mu_{N_{f}}^2+\zeta^2},\overbrace{\zeta,\ldots,\zeta}^{\nu})}{
{\cal Z}_{\nu}(\mu_1,\ldots,\mu_{N_{f}})} ~,
\eeq
\Rrule
\narrowtext
\noi
a more compact and convenient expression than the one provided in
Ref.\cite{NDW}. For example, in
the quenched case of $N_f=0$ it yields the simple relation
\beq
p_1^{(2)}(\zeta) =  \frac{\zeta}{2}\e^{-\zeta^2/4}
\det[I_{2+i-j}(\zeta)] ~,
\eeq
where the determinant is over a matrix of size $\nu\times \nu$. 
It is worth
pointing out that also the corresponding quenched distributions for
$\beta=1$ become exceedingly simple:
\beq
p_1^{(1)}(\zeta) =  {\rm const.}\, \zeta^{(3-\nu)/2}\e^{-\zeta^2/8}
\,{\mbox {\rm Pf}}\,[(i-j)I_{i+j+3}(\zeta)]~,
\label{beta1}
\eeq
where the indices $i$ and $j$ run between $-\nu/2+1$ and $\nu/2-1$, and the
Pfaffian is thus taken over a matrix of size $(\nu-1)\times (\nu-1)$
\cite{SmV}. 
For example, for $\nu=1$ this formula gives
\beq
p_1^{(1)}(\zeta) = \frac{\zeta}{4}\e^{-\zeta^2/8} ~,
\eeq
for $\nu=3$ we obtain
\beq
p_1^{(1)}(\zeta) = \frac{1}{2} \e^{-\zeta^2/8} I_3(\zeta) ~,
\eeq
while for $\nu=5$ it gives
\beq
p_1^{(1)}(\zeta) = \frac{1}{\zeta} \e^{-\zeta^2/8}
\left(3I_3(\zeta)^2 - 4I_2(\zeta)I_4(\zeta) + 
I_1(\zeta)I_5(\zeta)\right)~,
\eeq
all of these being correctly normalized.

By now, the effective
finite-volume partition functions are known in analytically closed forms for
$\beta=2$ \cite{JSV} and $\beta=1$ \cite{NN1}. 
They are also known for a general even number of pairwise degenerate
fermions for $\beta=4$ \cite{NN1} (this is a
technical restriction that we expect to be lifted eventually).
We shall here for convenience provide the precise detailed expressions
for all these known cases. In exhibiting explicit forms of
$\omega_k^{(\beta)}(\zeta_1,\ldots,\zeta_k; \{\m\})$
by substituting these expressions, we choose to set
$\nu=2/\beta-1$ in Eq.(\ref{Zeigenv}) and
introduce additional $\beta(\nu+1)/2-1$ massless flavors
instead.
The number of flavors
in this alternative picture, $N_f+\beta(\nu+1)/2-1$, will be
denoted by $n$. 
For the $\beta=4$ case,
we thus consider here only the case with an odd number of massless flavors,
so that the `effective'
number of massless flavors is even, $n\equiv 2a$.
We then get
\widetext
\Lrule
\bml
\bea
\omega_k^{(2)}(\zeta_1,\ldots,\zeta_k; \m_1,\ldots,\m_n)&=& {\rm 
const.}\,
 \e^{-\zeta_k^2/4} 
 \prod_{i=1}^k |\zeta_i|
 \frac{\prod_{j=1}^{k-1} (\zeta_k^2-\z^2_j)^2
 \prod_{l=1}^{n} (\zeta_k^2+\m^2_l)}{\prod_{i>j}^{k-1} 
 (\z_i^2-\z^2_j)^2
  \prod_{j=1}^{k-1}  \prod_{l=1}^{n} (\z_j^2+\m_l^2)
 }\,
\frac{\det[B^{(2)}]}{\det[A^{(2)}]}~,
\\
\omega_k^{(1)}(\zeta_1,\ldots,\zeta_k; \m_1,\ldots,\m_n)&=&{\rm 
const.}\,
 \e^{-\zeta_k^2/8} 
 \prod_{i=1}^k |\zeta_i|
\prod_{j=1}^{k-1} (\zeta_k^2-\z^2_j)
 \prod_{l=1}^{n} (\zeta_k^2+\m^2_l)\,
\frac{{\rm Pf}\,[B^{(1)}]}{ {\rm Pf}\,[A^{(1)}]}~,
\\
\omega_k^{(4)}(\zeta_1,\ldots,\zeta_k;
\m_1,\m_1,\ldots,\m_a,\m_a)&=& {\rm const.}\,
\e^{-\zeta_k^2/2} 
 \prod_{i=1}^k |\zeta_i|
 \prod_{j=1}^{k-1} (\zeta_k^2-\z^2_j)^4
 \prod_{l=1}^{a} (\zeta_k^2+\m^2_l)^2
\, \frac{{\rm Pf}\,[B^{(4)}]}{ {\rm Pf}\,[A^{(4)}]}~,
\eea
\eml
where the matrices $A^{(\beta)}$ and $B^{(\beta)}$
are given by 
($\tz_i\equiv \sqrt{\z_k^2-\z_i^2})$,
 $\tm_i\equiv \sqrt{\z_k^2+\m_i^2}$)
\bml
\bea
A^{(2)}&=&\left[\mu_i^{j-1} I_{j-1}(\mu_i)\right]_{i,j=1,\ldots,n}~ ,
\\
B^{(2)}&=&
\left[
\Bigl[\tm_i^{j-3} I_{j-3}(\tm_i)\Bigr]_{
i=1,\ldots,n \atop j=1,\ldots,n+2k-2}
\left[\tz_i^{j-3} I_{j-3}(\tz_i)\right]_{
i=1,\ldots,k-1 \atop j=1,\ldots,n+2k-2}
\left[\tz_i^{j-4} I_{j-4}(\tz_i)\right]_{
i=1,\ldots,k-1 \atop j=1,\ldots,n+2k-2}
\right] ~,
\\
A^{(1)}&=&
\left\{
\ba{ll}
\left[{D}_0(\mu_i,\mu_j)\right]_{i,j=1,\ldots,n}
& \ \ \ \ (n:\mbox{even}) ~, \\
\left[
\ba{ll}
 \left[{D}_0(\mu_i,\mu_j)\right]_{i,j=1,\ldots,n} &
 \left[{R}_0(\mu_j)\right]_{j=1,\ldots,n}\\
\left[-{R}_0(\mu_i)\right]_{i=1,\ldots,n} & 0
\ea
\right] & \ \ \ \ (n:\mbox{odd})~,
\ea
\right. 
\\
B^{(1)}&=&
\left\{
\ba{ll}
\left[
\ba{ll}
 \Bigl[{D}_1(\tm_i,\tm_j)\Bigr]_{i,j=1,\ldots,n} &
 \left[{D}_1(\tz_i,\tm_j)\right]_{i=1,..,k-1\atop j=1,\ldots,n}\\
 \left[{D}_1(\tm_i,\tz_j)\right]_{i=1,\ldots,n \atop j=1,\ldots,k-1} &
 \left[{D}_1(\tz_i,\tz_j)\right]_{i,j=1,\ldots,k-1}
\ea
\right]&\ \ \ \ 
(n+k: \mbox{odd})~,\\
\left[
\ba{lll}
 \Bigl[{D}_1(\tm_i,\tm_j)\Bigr]_{i,j=1,\ldots,n} &
 \left[{D}_1(\tz_i,\tm_j)\right]_{i=1,..,k-1\atop j=1,\ldots,n} &
 \Bigl[{R}_1(\tm_j)\Bigr]_{j=1,\ldots,n}\\
 \left[{D}_1(\tm_i,\tz_j)\right]_{i=1,\ldots,n \atop j=1,\ldots,k-1} &
 \left[{D}_1(\tz_i,\tz_j)\right]_{i,j=1,\ldots,k-1} &
 \left[{R}_1(\tz_j)\right]_{j=1,\ldots,k-1}\\
 \Bigl[-{R}_1(\tm_i)\Bigr]_{i=1,\ldots,n}&
 \left[-{R}_1(\tz_i)\right]_{i=1,\ldots,k-1}&
 0
\ea
\right] &  \ \ \ \ 
(n+k: \mbox{even})~,
\ea
\right. 
\\
&&{D}_\a(\z,\z')=
\int_0^1 dt\,t^2 \left(
\frac{I_{2\a}(t\z)}{\z^{2\a}}
\frac{I_{2\a+1}(t\z')}{\z'{}^{2\a+1}}-
\frac{I_{2\a+1}(t\z)}{\z^{2\a+1}}
\frac{I_{2\a}(t\z')}{\z'{}^{2\a}}\right),
\ \ \ 
{R}_\a(\z)= \frac{I_{2\a+1}(\z)}{\z^{2\a+1}}~,
\nonumber\\
A^{(4)}&=&
\left\{
\ba{ll}
\left[{I}_0(\mu_i,\mu_j)\right]_{i,j=1,\ldots,a}
& \ \ \ \ (a:\mbox{even})~, \\
\left[
\ba{ll}
 \left[{I}_0(\mu_i,\mu_j)\right]_{i,j=1,\ldots,a} &
 \left[{Q}_0(\mu_j)\right]_{j=1,\ldots,a}\\
\left[-{Q}_0(\mu_i)\right]_{i=1,\ldots,a} & 0
\ea
\right] &\ \ \ \   (a:\mbox{odd})~,
\ea
\right. 
\\
B^{(4)}&=&
\left\{
\ba{ll}
\!\left[
\ba{lll}
 \Bigl[{I}_4(\tm_i,\tm_j)\Bigr]_{i,j=1,\ldots,a} &
 \left[{I}_4(\tz_i,\tm_j)\right]_{i=1,..,k-1\atop j=1,\ldots,a} &
 \left[{S}_4(\tz_i,\tm_j)\right]_{i=1,..,k-1\atop j=1,\ldots,a} \\
 \left[{I}_4(\tm_i,\tz_j)\right]_{i=1,\ldots,a \atop j=1,\ldots,k-1} &
 \left[{I}_4(\tz_i,\tz_j)\right]_{i,j=1,\ldots,k-1} &
 \left[{S}_4(\tz_i,\tz_j)\right]_{i,j=1,..,k-1} \\
 \left[-{S}_4(\tm_i,\tz_j)\right]_{i=1,\ldots,a \atop j=1,\ldots,k-1} &
 \left[-{S}_4(\tz_i,\tz_j)\right]_{i,j=1,\ldots,k-1} &
 \left[{\overline{D}}_{4}(\tz_i,\tz_j)\right]_{i,j=1,..,k-1}
 \ea
\right]&\!\!\! (a: \mbox{even}),
\\
\!\left[
\ba{llll}
 \Bigl[{I}_4(\tm_i,\tm_j)\Bigr]_{i,j=1,\ldots,a} &
 \Bigl[-{Q}_4(\tm_i)\Bigr]_{i=1,\ldots,a} &
 \left[{I}_4(\tz_i,\tm_j)\right]_{i=1,..,k-1\atop j=1,\ldots,a} &
 \left[{S}_4(\tz_i,\tm_j)\right]_{i=1,..,k-1\atop j=1,\ldots,a} \\
 \Bigl[{Q}_4(\tm_j)\Bigr]_{j=1,\ldots,a} &
 0 &
 \left[{Q}_4(\tz_j)\right]_{j=1,\ldots,k-1} &
 \left[{O}_4(\tz_j)\right]_{j=1,\ldots,k-1} \\
 \left[{I}_4(\tm_i,\tz_j)\right]_{i=1,\ldots,a \atop j=1,\ldots,k-1} &
 \left[-{Q}_4(\tz_i)\right]_{i=1,\ldots,k-1} &
 \left[{I}_4(\tz_i,\tz_j)\right]_{i,j=1,\ldots,k-1} &
 \left[{S}_4(\tz_i,\tz_j)\right]_{i,j=1,..,k-1} \\
 \left[-{S}_4(\tm_i,\tz_j)\right]_{i=1,\ldots,a \atop j=1,\ldots,k-1} &
 \left[-{O}_4(\tz_i)\right]_{i=1,\ldots,k-1} &
 \left[-{S}_4(\tz_i,\tz_j)\right]_{i,j=1,\ldots,k-1} &
 \left[{\overline{D}}_{4}(\tz_i,\tz_j)\right]_{i,j=1,..,k-1}
 \ea
\right] &\!\!\! (a: \mbox{odd}),
\ea \right. \nonumber\\
&&\\
&&{I}_\a(\z,\z')=
\int_0^1 dt\,t \int_0^1 du
\left(
\frac{I_{\a-1}(2t\z)}{\z^{\a-1}}
\frac{I_{\a-1}(2tu\z')}{\z'{}^{\a-1}}-
\frac{I_{\a-1}(2tu\z)}{\z^{\a-1}}
\frac{I_{\a-1}(2t\z')}{\z'{}^{\a-1}}
\right),
\ \ \ 
{Q}_\a(\z)=\int_0^1 dt
\frac{I_{\a-1}(2t\z)}{\z^{\a-1}}~,
\nonumber\\
&&{S}_\a(\z,\z')=
\int_0^1 dt\,t^2 \int_0^1 du
\left(
\frac{I_{\a-1}(2t\z)}{\z^{\a-1}}
u \frac{I_{\a}(2tu\z')}{\z'{}^{\a}}-
\frac{I_{\a-1}(2tu\z)}{\z^{\a-1}}
\frac{I_{\a}(2t\z')}{\z'{}^{\a}}
\right),
\ \ \ {O}_\a(\z)=\int_0^1 dt\,t
\frac{I_{\a}(2t\z)}{\z^{\a}} ~,
\nonumber\\
&&\overline{D}_\a(\z,\z')=
\int_0^1 dt\,t^3 \int_0^1 du\,u
\left(
\frac{I_{\a}(2t\z)}{\z^{\a}}
\frac{I_{\a}(2tu\z')}{\z'{}^{\a}}-
\frac{I_{\a}(2tu\z)}{\z^{\a}}
\frac{I_{\a}(2t\z')}{\z'{}^{\a}}
\right) ~.
\nn
\eea 
\eml
\Rrule
\narrowtext
\noi
In all cases the normalization constants are fixed uniquely by the
requirement that the probabilities sum up to unity.

By construction, the individual distributions 
${p}^{(\beta)}_{k}(\zeta;\{\mu\})$
sum up to the microscopic spectral density $\rho^{(\beta)}_S(\zeta;\{\mu\})$:
\beq
\rho^{(\beta)}_S(\zeta;\{\mu\})=\sum_{k=1}^\infty
{p}^{(\beta)}_{k}(\zeta;\{\mu\})~.
\eeq
To illustrate this, we plot in Fig.1
${p}^{(2)}_{k}(\zeta)$ for $k=1,\ldots,4$, their sum, and
$
\rho^{(2)}_S(\zeta)=|\zeta| \Bigl(J_0^2(\zeta)+J_1^2(\zeta)\Bigr)/2
$
for the quenched ($N_f=0$) chiral unitary ensemble with $\nu=0$. One
clearly sees how the microscopical spectral density gradually
builds up from the individual eigenvalue distributions.

We finally turn to the issue of universality.
It was proven in Refs.\cite{SeV,Wid} that the diagonal
elements of the quaternion kernels
$K^{(1,4)}(\z,\z';\{\m\})$
for the orthogonal and symplectic ensembles can be
constructed from the scalar kernel
$K^{(2)}(\z,\z';\{\m\})$ of a unitary ensemble
with a related weight function.
As the scalar kernel in the microscopic limit
(\ref{microlim})
is insensitive to the details of the potential $v(x)$
either in the absence \cite{ADMN,KF} or
in the presence of finite and non-zero masses $\mu$ \cite{DN98a},
so are the corresponding quaternion kernels. Furthermore,
${p}^{(\beta)}_k(\zeta;\{\m\})$ can be expressed in terms of
the Fredholm determinant of $K^{(\beta)}(\z,\z';\{\m\})$
\cite{Meh,TW93}:
\beq
{p}^{(\beta)}_k(\zeta;\{\m\})=
\frac{(-1)^k}{(k-1)!}\frac{\partial}{\partial \zeta}
\left.
\Bigl(\frac{\partial}{\partial t}\Bigr)^{k-1}
\det (1-t \,\hat{K}_\zeta^{(\beta)})
\right|_{t=1} ~.
\eeq
Here $\hat{K}_\zeta^{(\beta)}$ stand for the integral operators
with convoluting kernels
$K^{(\beta)}(\varsigma,\varsigma';\{\m\})$
over an interval $0\leq \varsigma,\varsigma'\leq \zeta$.
The universality of the probability distribution
${p}^{(\beta)}_k(\zeta;\{\m\})$ is hence guaranteed. It is nevertheless
instructive
to see how this manifests itself in our explicit computation.
For a generic Random Matrix Theory potential
the exponential factor $\e^{-\beta (N-k) x_k}$
which is produced by the shift $x_i \to x_i +x_k$
is simply replaced by $\exp\bigl(-({\beta}/{8})
\bigl(\pi\rho(0)\bigr)^2 x_k
\bigl(1+{\cal O}(1/N) \bigr)\bigr)$,
and thus yields an identical factor $\e^{-\beta \z_k^2/8}$
in the microscopic limit (\ref{microlim}) \cite{NDW}. Based on the
universality theorems in Refs.\cite{ADMN,SeV} one readily establishes that
also the remaining ratio of partition functions, and in particular the
full expression (\ref{bboxOmega}), is universal. 

PHD would like to thank the Institute for Nuclear Theory
at the University of Washington for its hospitality and
DOE for partial support during the completion of this work.
We thank R. Niclasen for help with the figure preparation.
The work of PHD was supported in part by EU TMR Grant no.\ 
ERBFMRXCT97-0122
and the work of SMN was supported in part by JSPS, and
by Grant-in-Aid no.\ 411044
from the Ministry of Education, Science, and Culture, Japan.

\widetext
\Lrule
\begin{figure}
\epsfxsize=320pt
  \begin{center}
    \leavevmode
\epsfbox{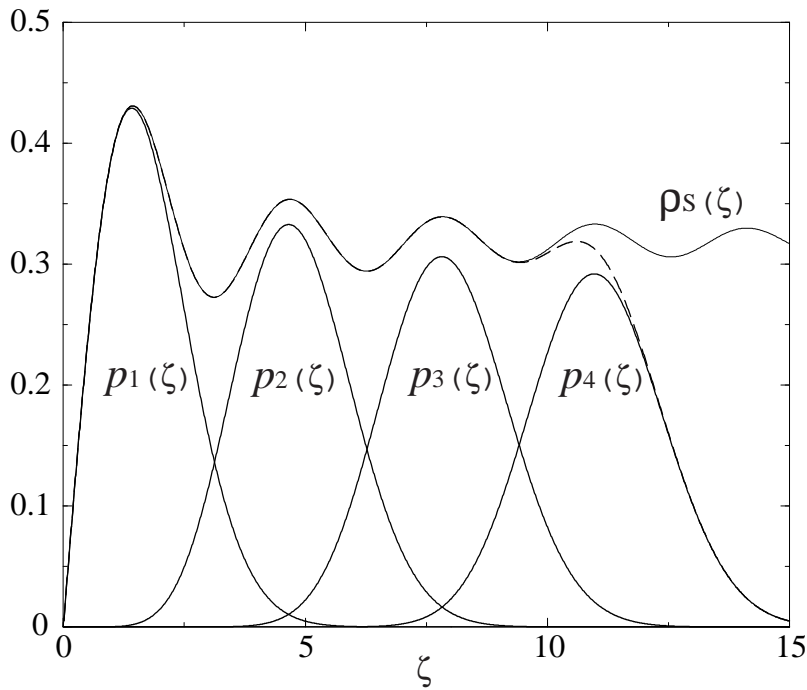} 
  \end{center}
\caption{
Microscopic spectral density
$\rho_S(\zeta)$ (thin line),
$k$-th smallest eigenvalue distribution
$p_k(\zeta)$, $k=1, 2, 3, 4$ (solid lines), and their sum
$\sum_{k=1}^4 p_k(\zeta)$ (broken line)
for the quenched chiral unitary ensemble ($\beta=2, N_f=\nu=0$).
}
\end{figure}

\begin{references}
\bibitem{LS}
H. Leutwyler and A. Smilga,
Phys. Rev. {\bf D46}, 5607 (1992).
\bibitem{SmV}A. Smilga and J.J.M. Verbaarschot,
Phys. Rev. {\bf D51}, 829 (1995).
\bibitem{ShV}
E.V. Shuryak and J.J.M. Verbaarschot,
Nucl. Phys. {\bf A560}, 306 (1993).
\bibitem{V}J.J.M. Verbaarschot and I. Zahed,
Phys. Rev. Lett. {\bf 70}, 3852 (1993);
\newline J.J.M. Verbaarschot, Nucl. Phys. {\bf B426} (1994) 559;
Phys. Rev. Lett. {\bf 72}, 2531 (1994).
\bibitem{HV}M.A. Halasz and J.J.M. Verbaarschot,
Phys. Rev. {\bf D52}, 2563 (1995).
\bibitem{VW}
For an exhaustive list of references, see:
J.J.M. Verbaarschot and T. Wettig,  hep-ph/0003017.
\bibitem{D}P.H. Damgaard, Phys. Lett. {\bf B424}, 322 (1998).
\newline
G. Akemann and P.H. Damgaard, Nucl. Phys. {\bf B528}, 411 (1998);
Phys. Lett. {\bf B432}, 390 (1998);  hep-th/9910190.
\bibitem{OTV}
J.C. Osborn, D. Toublan and J.J.M. Verbaarschot,
Nucl. Phys. 
{\bf B540}, 317 (1999);
\newline 
P.H. Damgaard, J.C. Osborn, D. Toublan and J.J.M. Verbaarschot,
Nucl. Phys. {\bf B547}, 305 (1999);
\newline  
D. Toublan and J.J.M. Verbaarschot,
Nucl. Phys. {\bf B560}, 259 (1999).
\bibitem{NS}T.~Nagao and K.~Slevin,
J. Math. Phys. {\bf 34}, 2075 (1993).
\bibitem{For1}
P.J. Forrester, Nucl. Phys. {\bf B402}, 709 (1993).
\bibitem{For2}
P.J. Forrester, J. Math. Phys. {\bf 35}, 2539 (1994).
\bibitem{AST}
A.V. Andreev, B.D. Simons, and N. Taniguchi,
Nucl. Phys. {\bf B432}, 487 (1994).
\bibitem{NF95}T. Nagao and P.J. Forrester,
Nucl. Phys. {\bf B435}, 401 (1995).
\bibitem{DN98a}
P.H. Damgaard and S.M. Nishigaki,
Nucl. Phys. {\bf B518}, 495 (1998).
\bibitem{WGW}
T. Wilke, T. Guhr, and T. Wettig,
Phys. Rev. {\bf D57}, 6486 (1998).
\bibitem{NN1}
T. Nagao and S.M. Nishigaki,  hep-th/0001137,
Phys. Rev. {\bf D62} (2000) in press.
\bibitem{AK} 
G. Akemann and E. Kanzieper,  hep-th/0001188.
\bibitem{NN2}
T. Nagao and S.M. Nishigaki,  hep-th/0003009,
Phys. Rev. {\bf D62} (2000) in press.
\bibitem{SlN}
K. Slevin and T. Nagao,
Phys. Rev. Lett. {\bf 70}, 635 (1993).
\bibitem{TW}
C.A. Tracy and H. Widom,
Commun. Math. Phys. {\bf 161}, 289 (1994).
\bibitem{FH}
P.J. Forrester and T.D. Hughes,
J. Math. Phys. {\bf 35}, 6736 (1994).
\bibitem{NF98}T. Nagao and P.J. Forrester,
Nucl. Phys. {\bf B509}, 561 (1998).
\bibitem{NDW}
S.M. Nishigaki, P.H. Damgaard, and T. Wettig,
Phys. Rev. {\bf D58}, 087704 (1998).
\bibitem{ADMN}
S. Nishigaki, Phys. Lett. {\bf B387}, 139 (1996);
\newline
G. Akemann, P.H. Damgaard, U. Magnea, and S. Nishigaki,
Nucl. Phys. {\bf B487}, 721 (1997).
\bibitem{KF}
E. Kanzieper and V. Freilikher,
Philos. Mag. {\bf B77}, 1161 (1998).
\bibitem{SeV}
M.K. \c{S}ener and J.J.M. Verbaarschot, Phys. Rev. Lett. {\bf 81},
248 (1998).\newline
B. Klein and J.J.M. Verbaarschot,  hep-th/0004119.
\bibitem{Wid}
H. Widom, J. Stat. Phys. {\bf 94}, 347 (1999).
\bibitem{JSV}R. Brower, P. Rossi, and C.-I. Tan, Nucl. Phys. {\bf B190},
699 (1981).\newline  
A.D. Jackson, M.K. \c{S}ener, and J.J.M. Verbaarschot,
Phys. Lett. {\bf B387}, 355 (1996).
\bibitem{Meh}
M.L. Mehta, 
{\it Random Matrices}, 2nd Ed. (Academic, San Diego, 1991).
\bibitem{TW93}
C.A. Tracy and H. Widom,
Springer Lecture Note in Physics {\bf 424}, 103 (1993).
\end{references}
\end{document}